\documentclass[a4paper,fleqn,usenatbib,referee,epsfig,color]{mnras}

\usepackage[T1]{fontenc}
\usepackage{ae,aecompl}

\usepackage{graphicx}	
\usepackage{amsmath}	
\usepackage{amssymb}	

\newif\ifAMStwofonts
\newcommand{\be}{\begin{equation}}
\newcommand{\ee}{\end{equation}}
\newcommand{\bea}{\begin{eqnarray}}
\newcommand{\eea}{\end{eqnarray}}

\title[Halo Magnetic field Structure]{\bf  Exact axially symmetric galactic dynamos }
\author[R.N. Henriksen, A. Woodfinden, J.A. Irwin]
{R. N. Henriksen$^1$\thanks{henriksn@astro.queensu.ca},
A. Woodfinden$^1$\thanks{17aw14@queensu.ca}, and J. A. Irwin$^1$\thanks{irwinja@queensu.ca}\\
$^1$Dept. of Physics, Engineering Physics \& Astronomy, Queen's University, Kingston, Ontario, K7L 3N6, Canada\\}
\date{Accepted XXX. Received YYY; in original form ZZZ}
\pubyear{2017}
\begin{document}
\label{firstpage}
\pagerange{\pageref{firstpage}--\pageref{lastpage}}
\maketitle

\begin{abstract}
  We give a selection of exact dynamos in axial symmetry on a galactic scale. These include some steady examples, at least one of which is wholly analytic  in terms of simple functions and has been discussed elsewhere. Most solutions are found in terms of special functions, such as associated Lagrange or hypergeometric functions. They may be considered exact in the sense that they are known to any desired accuracy in principle. The new aspect  developed here is to present  scale invariant solutions  with zero resistivity that are self-similar in time.  The time dependence is either a power law or an exponential  factor, but since the geometry of the solution is self-similar in time we do not need to fix a time to study it. Several examples are discussed.  Our results demonstrate (without the need to invoke any other mechanisms) X-shaped magnetic fields and (axially symmetric) magnetic spiral arms (both of which are well observed and documented) and predict reversing rotation measures in galaxy halos (now observed in the CHANG-ES sample) as well as the fact that planar magnetic spirals are lifted into the galactic halo.  

\end{abstract}
\section{Introduction}
In our previous work on the classical galactic dynamo \cite{Hen2017}, \cite{Hen2017b}, 
we have studied the classical \cite{M1978} steady dynamo equations using the assumption of scale invariance.  Because of the modal ans\"atz employed, we were constrained to study the magnetic field lying on cones at relatively small angles to the galactic plane. These fields show and confirm  some  of the behaviours found by previous workers (see e.g \citet{B2014}, \citet{Beck2015}, \citet{M2015}, \citet{KF2015} for introductory references). These include `X type' fields (\cite{Kr2015}) and `parity inversion' (i.e. sign change in the azimuthal field in axial symmetry above the galactic disc \cite{SS1990}, \cite{BDMSST92}, \cite{MS2008}). Moreover in the spiral mode study \cite{Hen2017b}, we have made a clear prediction that the planar magnetic spirals observed in face-on galaxies \cite{Beck2015} are `lifted' into the galactic halo.

We have also predicted Faraday rotation screens that are compatible with new data arriving from
  the CHANG-ES\footnote{Continuum Halos in Nearby Galaxies -- an EVLA Survey \citep{WI2015}.} consortium (e.g. \cite{SPK2016}, \cite{CMP2016}).  However, our previous limitation to small angle cones has led to questions about the global behaviour of the halo magnetic fields. In this paper we  remove the limitation of small angle cones  for axially symmetric galactic dynamos by treating both steady and time dependent  'exact' solutions and show that their forms confirm the conclusions of the earlier work.

Our work is all in the context of the unmodified classical theory except for the assumption of scale invariance. The latter symmetry frequently develops in complex systems well away from boundary conditions \cite{Hen2015}, \cite{Barenblatt96}, such as may be the major part of a galactic disc. The earlier work assumed that a steady state had been reached so that the origin and evolution of the magnetic field was not discussed. We  mitigate this somewhat in this paper by finding solutions that are self-similar in time. That is, the time dependence is either a power  or exponential  temporal factor, so that the spatial behaviour does not change to within a time dependent global factor. This will allow eventual contact between a `seed' galactic field and present field strength, but that is not the prime concern of this paper.

\section{Exact, Steady, Scale Invariant, Dynamo Fields }

We refer to \cite{Hen2017} for the formulation of the scale invariant equations describing classical, steady, axially symmetric dynamo magnetic fields. {The basic equation for the dynamo vector potential ${\bf A}$ is, in the absence of electrostatic fields, 
\be
0={\bf v}\wedge \nabla\wedge {\bf A}-\eta\nabla\wedge\nabla\wedge {\bf A}+\alpha_d\nabla\wedge {\bf A}.\label{eq:Afield}
\ee
Only the curl of the vector potential appears in these equations so that they may also be written in terms of the magnetic field  as in \cite{Hen2017}. }

The equations that we will solve in this section deal directly with the magnetic field.  Unlike a solution in terms of the vector potential, the divergence constraint must be applied separately. In either formulation, the equations for {\it divergence-free solutions are over determined  in the self-similar steady state}, unless other restrictions are applied. 

The physical magnetic field is written  conveniently throughout this series as 
\be
{\bf B} \equiv {\bf b}\sqrt{4\pi\rho}.\label{eq:Bfield}
\ee
In this expression $\rho$ is a completely arbitrary constant mass density, so that numerically $\sqrt{4\pi\rho}$ is simply a constant that may be absorbed into multiplicative constants, which appear subsequently in the solutions. It serves solely to convert the Dimensions of the magnetic field ${\bf b}$ to those of a velocity, so that we need not introduce electromagnetic Units into our argument. If we set $4\pi\rho=1$ in cgs Units, then ${\bf b}$ may be expressed in Gauss.

The assumption of scale invariance in the axially symmetric {\it steady state} dynamo requires that { (this is explained more fully in equations (4), (5) and (9) of \cite{Hen2017})}
\be
{\bf b}=(\delta r)^{(1-a)}\bar{\bf b}(\zeta), \label{eq:bfield}
\ee
while the velocity field that appears in the dynamo equations takes the form
\be
{\bf v}=(\bar \eta(\zeta) \delta)(\delta r)^{(1-a)}(u,v,w).\label{eq:vfield}
\ee
The scale invariant variable $\zeta$ is 
\be 
\zeta=\frac{z}{r}.\label{eq:zeta}
\ee 
The quantity $\zeta$ is therefore constant on those cones with vertex at the galactic centre and  having semi vertex angle $\pi/2-arctan(z/r)$. 
We have indicated in equation (\ref{eq:vfield})  that the barred quantity (the amplitude of the resistive diffusion-see below) may vary on cones. We ignore this possibility in this article, but it is a possible future generalization.

In these equations $r$ is the cylindrical radius of a point ( $\{r,\phi,z\}$ relative to the minor galactic axis with origin at the centre) in the galaxy  disc and halo, and the  {\it scaled }  cylindrical velocity components (taken constant) are labelled $(u,v,w)$ in Units of $\bar\eta\delta$. The quantity $\delta$ may be thought of as a convenient reciprocal spatial scale and the resistive diffusion coefficient $\eta$ is given by 
\be
\eta =\bar\eta(\zeta)(\delta r)^{(2-a)}.\label{eq:diffusion}
\ee
Hence the velocity is measured in Units of a diffusion velocity at the scale $1/\delta$. 

Similarly the {\it sub scale} dynamo  due to helically turbulent velocity (the familiar $\alpha$ parameter, which is here labelled $\alpha_d$) is given by the assumption of scale invariance in the form 
\be
\alpha_d  =(\delta r)^{(1-a)}\bar\alpha_d(\zeta),
\ee
where the  indicated possible dependence of $\bar\alpha_d$ on $\zeta$ will be ignored in what follows.

The parameter $a$ is a pure number lying practically in the range $[0,3]$ that defines the similarity `class' \cite{CH1991}. It is the ratio of spatial ($L$) to temporal ($T$) powers that occur in the Dimensions of a hypothetical global constant  ($a\equiv \alpha/\delta$) that may  govern the scale invariance \cite{Hen2015}. For example it would be $1$ for a globally constant speed, $2$ for a global 
specific angular momentum and $3$ for globally conserved magnetic flux.  When $\alpha=0$ there is a quantity solely with the Dimensions of time (such as an angular velocity) that is constant globally. However for our purposes  in this study $a$ is best regarded as an arbitrary parameter, since we do not know the globally conserved quantity in advance. If a distinction is made between the vertical and radial characteristic scales, slightly more general radial dependences are permitted \cite{Hen2017}.  As a guide, Table~\ref{table:a} provides a possible physical identification of the similarity class.

  \begin{table*}
    \centering
    \begin{minipage}{110mm}
 \caption{Similarity Class Identification$^{i,\,ii}$\label{table:a}}
 \begin{tabular}{@{}lcc@{}}
   \hline
       {a} & {Dimensions of X} & {Possible Identification}\\
       \hline
 0    & $T^q$         & Angular velocity if $q\,=\,-1$\\
 1    & $L^n/T^n$     & Linear velocity if $n\,=\,1$\\
 3/2  & $L^{3n}/T^{2n}$& Keplerian orbits if $n\,=\,1$\\
 2    & $L^{2n}/T^n$   & Specific angular momentum if $n\,=\,1$\\
 3    & $L^{3n}/T^n$   & Magnetic flux if $n\,=\,1$\\
 \hline
 \end{tabular}
 \\
      {\it i} Recall that magnetic field and velocity have the same Dimensions when the field\\is divided by the square root of an arbitrary density.\\
     {\it ii} Recall that, generally, ${\rm a}\,\equiv\,\alpha/\delta\,=\,p/q$, where the globally\\ conserved quantity, X, has Dimensions [X] = $L^p/T^q$
\end{minipage}
  \end{table*}

{The magnetic field equations are given below  (\ref{eq:beqs}), and they follow from equation (\ref{eq:Afield}) when the self-similar forms are used for $\bf A$ and ${\bf b}$. One should note that the same equations appeared in \cite{Hen2017}, but $Z$ was used in place of $\zeta$.} Thus with axial symmetry and scale invariance we have 

\bea
0&=& \bar b_r\Delta+ v\bar b_z-w\bar b_\phi+\bar b'_\phi,\nonumber\\
0&=& -(1+\zeta^2)\bar b'_r+ w\bar b_r- u\bar b_z+\bar b_\phi\Delta+(1-a)\bar b_z+(2-a)\zeta\bar b_r,\label{eq:beqs}\\
0&=& \zeta\bar b'_\phi-(2-a)\bar b_\phi+ u\bar b_\phi- v\bar b_r+\bar b_z\Delta.\nonumber
\eea
 The divergence-free condition adds the constraint 
 \be
 (2-a)\bar b_r-\zeta\bar b'_r+\bar b'_z=0.\label{eq:divb}
 \ee

In these equations the prime indicates $d/d\zeta$, and only the  similarity class  $a$ and 
\be
\Delta\equiv \frac{\bar\alpha_d}{\bar \eta\delta} \label{eq:dynamonum}
\ee
are parameters.
The latter is a Reynolds number of the sub scale turbulence, which is effectively the dynamo number \cite{B2014}. We treat this as a constant in the following analysis, which requires $\bar\eta(\zeta)$ and $\bar\alpha_d(\zeta)$ either to be constant or to be proportional to the same function.  If $\Delta$ is allowed to vary on cones, the over determinedness is lifted but the equations become strongly non-linear. We proceed in the following sub-sections to look at  special cases. \footnote{  We have MAPLE scripts of these examples that 
  we intend to post on a web page in time for the publication of this article. The address of the web page will be available from henriksn@astro.queensu.ca.} 

When considering exact solutions, it should be remembered that they do not know about the galactic disc `a priori'. In particular should outflow be present, the solution is not guaranteed to have continuity across the disc when the sign of the outflow is reversed (as is required unless there is a `wind' perpendicular to and traversing the disc,  as might be the case for a galaxy orbiting in a cluster environment). In such cases the boundary conditions must be chosen so that the vertical magnetic field is continuous (thin disc)  and  one of the tangential components is either symmetric or anti-symmetric. 

\subsection{An Alpha/Alpha Dynamo}

In this example we allow arbitrary $a$ but set $v=w=0$.  Moreover $u=-1$ so there is an inward `migration'  of the  mean gas flow. Then equations (\ref{eq:beqs}) give 
\bea
\bar b_r&=&-\frac{\bar b'_\phi}{\Delta},\nonumber \\
\bar b_z&=& \frac{(3-a)\bar b_\phi-\zeta\bar b'_\phi}{\Delta},\label{eq:alpha/alpha}\\
0&=& (1+\zeta^2)\bar b''_\phi-2(2-a)\zeta \bar b'_\phi+(\Delta^2+(2-a)(3-a))\bar b_\phi.\nonumber
\eea
The first two equations of this set show that the constraint (\ref{eq:divb}) is identically satisfied.  

The equation for $\bar b_\phi$ may be solved formally in terms of associated Legendre functions (or hypergeometric series) in the form 
\be
\bar b_\phi=(1+\zeta^2)^{\frac{3-a}{2}}\big [C_1 P^\mu_\nu(i\zeta)+C_2 Q^\mu_\nu(i\zeta)\big ],\label{eq:barbphi}
\ee
where 
\be
\nu\equiv \frac{\sqrt{1-4\Delta^2}-1}{2},~~~~~~~~\mu\equiv a-3.
\ee 
These functions are well known to various mathematical libraries, although the treatment of the cut along the real axis should be chosen so as to ensure continuity across the disc along the real axis. Frequently  one can only construct a solution that is similar above and below the disc by `reflection' through the plane (changing the sign to maintain the continuity of the vertical component).  This places the source of the magnetic and velocity fields in the disc. Other forms of the solutions can be relevant if for example there is a `wind' blowing through the galaxy.

\begin{figure}
\begin{tabular}{cc} 
\rotatebox{0}{\scalebox{0.5} 
{\includegraphics{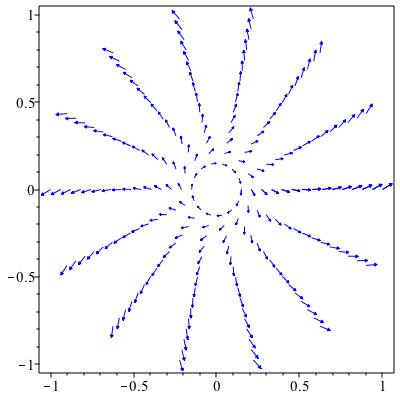}}}&
\rotatebox{0}{\scalebox{0.5} 
{\includegraphics{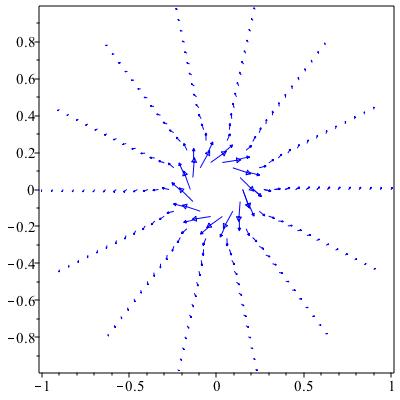}}}\\
{\rotatebox{0}{\scalebox{0.5} 
{\includegraphics{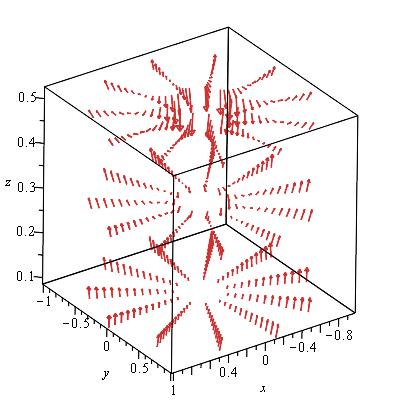}}}}&
\rotatebox{0}{\scalebox{0.5} 
{\includegraphics{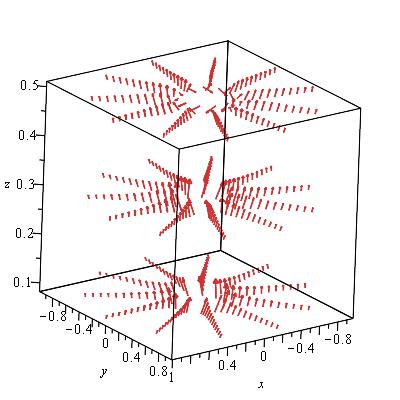}}}
\end{tabular}
\caption{At upper left we show a cut at $z=0.25$ for $0.15\le r\le 1$ and parameter set $\{a,\Delta,C1,C2\}$=$\{0,2.0,1,0\}$. At upper right we show the same cut for the same parameters when $a=2$. At lower left the three dimensional structure for $a=0$ and otherwise the same parameters is shown for $0.25\le r\le 1$ and $0.2\le z\le 0.5$. At lower right we have the same structure over the same range with the same parameters when $a=2$.}    
\label{fig:steadyu-1}
\end{figure}

When $a=3$, that is a flux conserved case,  the general case of this section becomes the example discussed in \cite{Hen2017} (equation (31) of that paper). Figure 2 of that paper illustrates some properties of the solution, notably the parity inversion of the magnetic field and the development of `X type behaviour only with height. It should be noted that such an example does not exhaust the  behaviour possible, since these images represent a rather specific choice of parameters.
In figure(\ref{fig:steadyu-1}) of this paper we illustrate the effect of varying $a$ on the global magnetic field.

At upper left of figure  (\ref{fig:steadyu-1}) we show a cut at $z=0.25$ with $0.15\le r \le 1$ for the similarity class $a=0$. One sees a clear sign change in what would be the RM screen with radius. In this class the strength of the field increases with radius as $r$.  One sees axially symmetric magnetic spirals. On the upper right of the figure we show the same cut for the similarity class $a=2$. A major difference is that now the field decreases with radius as $1/r$,  but in addition there is little visible evidence for spiralling magnetic structure at small radius. It is present  at large radius however.

At lower left we show the three dimensional structure of the $a=0$ class. Here $0.25\le r \le 1$  and $0.1\le z \le 0.5$. Sign reversals  that appear between large and small radius and large and small heights are evident. The apparent increase with height at small radius is actually an increase with increasing $\zeta$. This value is $\zeta=2$ at $z=0.5$ and $r=0.25$ . This corresponds to an angle with the galactic axis of only $26^\circ.6$ , so that the solution actually predicts increasing strength and more vertical direction near the galactic axis. At lower right the three dimensional structure for the class  $a=2$ is shown. Other parameters and coordinate ranges are the same. In all cases $\Delta=2.0$.  The major difference with the class $a=0$ is that near the axis of the galaxy (i.e. large $\zeta$)  the increasing field component is more azimuthal. 

We recall that the case $a=0$ implies a constant time Unit such as an angular velocity (or a jump in angular velocity-there could be an arbitrary length Dimension) while the $a=2$ class implies a constant angular momentum.  The $a=1$ class implies a constant velocity and the only radial dependence is through the definition of $\zeta$. We find that a similar cut to that shown in figure (\ref{fig:steadyu-1}) for the class $a=0$ is very similar for $a=1$, except that  the inner structure  in that case shows more spiral structure. A higher cut in $z$ reveals a polarization magnetic ring inside a magnetic ring that is much as seen in the $a=0$ case at larger $z$.

\subsection{A Classic Dynamo with Diffusion}

In the previous section the sub-scale turbulent helicity was the dominant source of the steady magnetic field. In this section we pursue the opposite extreme where there is no sub-scale term ($\Delta=0$). There is nevertheless dynamo action that produces a steady, axially symmetric, magnetic field through distortion by a mean velocity field (e.g. \cite{HI2016}), which field is ultimately `quenched' by resistive diffusion. This is an example of a  $\eta-\omega$ dynamo.

When $\Delta=0$ equations (\ref{eq:beqs}) require ($v\ne 0$)
\bea
\bar b_r&=&\frac{\zeta \bar b'_\phi}{v}+\frac{\bar b_\phi(u-(2-a))}{v}\nonumber\\
\bar b_z&=&\frac{w\bar b_\phi-\bar b'_\phi}{v},\label{eq:Dvel}\\
0&=& -(1+\zeta^2)\bar b''_\phi+\big (\zeta(2(2-a)-(1+u))+ w\big )\bar b'_\phi+\big ((2-a)(u-(2-a))-\frac{w}{\zeta}\big )\bar b_\phi.\nonumber
\eea
However the divergence constraint (\ref{eq:divb}) requires
\be
0= -(1+\zeta^2)\bar b''_\phi+\big (\zeta(2(2-a)-(1+u))+ w\big )\bar b'_\phi+\big ((2-a)(u-(2-a))\big )\bar b_\phi,\label{eq:divb2}
\ee
and this is only guaranteed to hold by a solution of equation (\ref{eq:Dvel}) if $w=0$ (by inspection). Of course approximations may be found by using the divergence condition and making $w/\zeta$ small compared to other terms in the factor multiplying $\bar b_\phi$, but we will proceed with $w=0$ so that there is neither wind nor accretion.\footnote{ One may note that were the factor $(1-a)$ multiplying $\bar b_z$ in the second of equations (\ref{eq:beqs}) equal instead to $(2-a)$ then the dynamo equation and the divergence condition would agree exactly. However all of our checks have served to confirm the factor $(1-a)$.} 

Taking $w=0$, the solution to equation ((\ref{eq:divb2}) -equivalently the last of equations (\ref{eq:Dvel})) may be written as 
\be
\bar b_\phi=(1+\zeta^2)^{-\mu/2}\big (C1~P^\mu_\nu(i\zeta)+C2~Q^\mu_\nu(i\zeta)\big),\label{eq:D+vel}
\ee
where 
\be
\nu\equiv \frac{u-1}{2},~~~~~~\mu \equiv \frac{u+2a-5}{2}.
\ee
The azimuthal velocity enters only in the amplitude of the radial and vertical magnetic field components according to equation (\ref{eq:Dvel}).

This result is a generalization of the first example given in \cite{Hen2017} (equation  (24) of that paper), which example is found by setting $a=2$ and $u=0$ in equations (\ref{eq:Dvel}) above. That case is defined by the remarkably simple analytic expression for $\bar b_\phi $ 
\be
  \bar b_\phi =\bar b_\phi(0)+C\ln{|\zeta+sgn(\zeta)\sqrt{1+\zeta^2}|}=\bar b_\phi(0)+C sinh^{-1}(\zeta)
 \ee
where the first expression is explicitly valid for both sides of the disc.  The derivative equal to 
\be
\bar b'_\phi=\frac{C}{\sqrt{1+\zeta^2}}= C\frac{r}{\sqrt{r^2+z^2}},
\ee
allows the poloidal field components to be readily found from equations (\ref{eq:Dvel}). We observe that with $C$ continuous across the plane, both $\bar b_z$ and $\bar b'_\phi$ are continuous. Equation (\ref{eq:Dvel}) implies that $\bar b_r$ will change sign on crossing the plane.  In order to maintain symmetry about the plane  (i.e. a symmetrical action of the sources in the galactic disc) we should therefore set $\bar b_\phi(0)=0$ and so allow $\bar b_\phi$ also to be anti-symmetric about the plane.  Changing the sign of $C$ on crossing the plane is not permitted since $\bar b_z$ must be continuous.

One important property of this  simple example is that {\it it contains the 'X type' magnetic field structure}. Thus the angle  of the field line with the plane,$\lambda$, is given by 
\be
\lambda =\arctan{\frac{\bar b_z}{\bar b_r}}=-\frac{1}{\zeta},
\ee
so that 
\be
tan(\pi/2-\lambda)=\zeta.
\ee
This shows the poloidal magnetic field lines to lie on cones whose generators project onto the sky plane as straight lines. 
{To untangle `X type' field structure based on synchrotron emission, a model for the distribution of the relativistic electrons as well as for the magnetic field is required.  In an edge-on galaxy with axial symmetry  in both field and density we might expect the emission to reflect somewhat the tangent point to a circle of given radius.  This is because this point is closest to the centre of the galaxy and the field and electron density may both be expected to decline with radius.  }

\begin{figure}
\begin{tabular}{cc} 
\rotatebox{0}{\scalebox{0.3} 
{\includegraphics{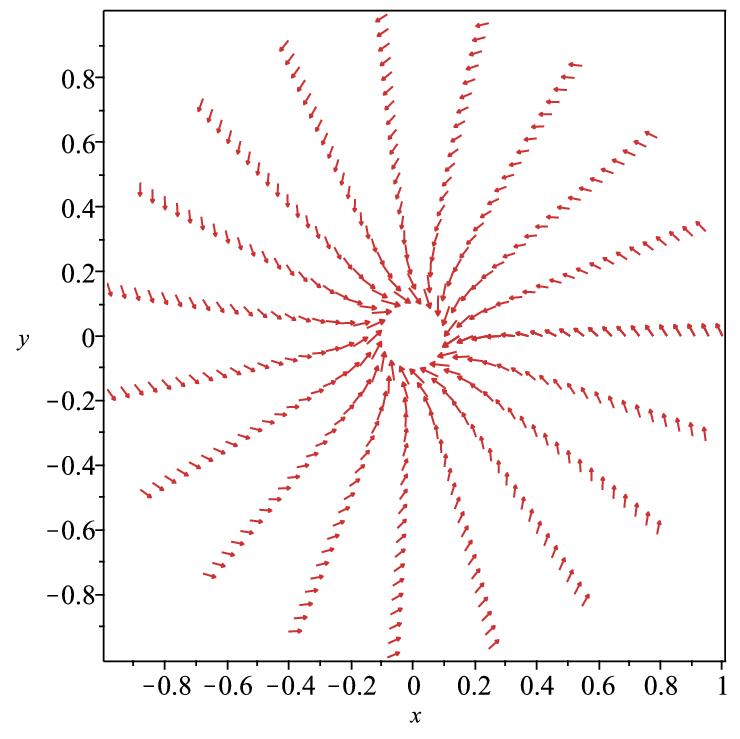}}}&
\rotatebox{0}{\scalebox{0.65} 
{\includegraphics{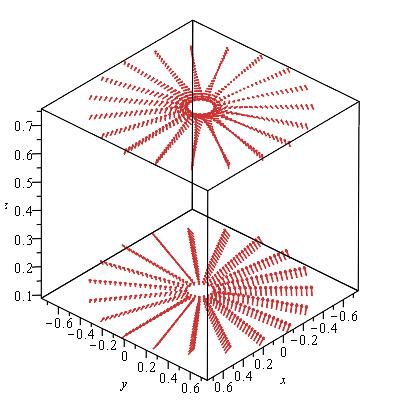}}}\\
{\rotatebox{0}{\scalebox{0.6} 
{\includegraphics{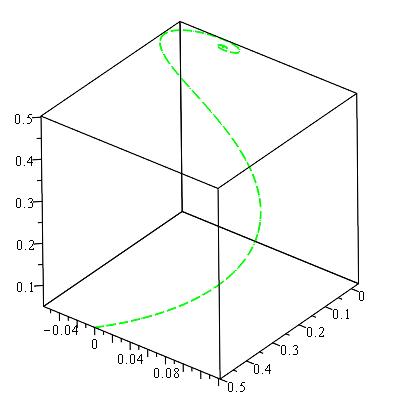}}}}&
\rotatebox{0}{\scalebox{0.5} 
{\includegraphics{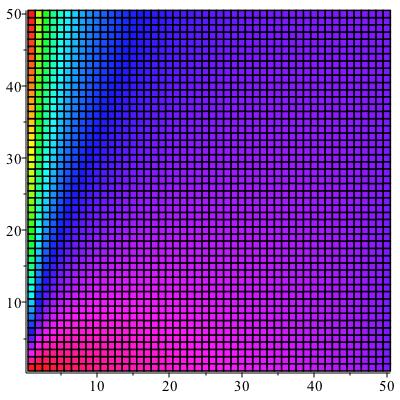}}}
\end{tabular}
\caption{At upper left we show a cut through the pure $\eta-\omega$ exact solenoidal solution at $z=0.35$. We have set $C=-\bar b_\phi(0)=-1$. The radius runs from $0.15$ to $1.0$. At upper right there is a 3D plot of the same solution wherein each of $r$ and $z$ run from $0.15$ to $0.75$.  At lower left we show a field line from the same solution that originates in the disc at  $r=0.5$, $\phi=0$ and $z=0.05$ and continues to positive $z$. It spirals into and crosses $r=0$ near $z=0.5$ and is then reproduced on the other side of the axis, forming a closed twisted loop. At lower right we show the Faraday screen rotation measure in the first quadrant for this solution. We take $\bar u_\phi=1$ in all images except at upper right where $\bar u_\phi=2$, in orde to emphasixe the azimuthal field near the axis.   }
\label{fig:analytvphi}
\end{figure}

In figure 1 of \cite{Hen2017} some properties of this solution are illustrated.  This figure is copied below for reference in figure (\ref{fig:analytvphi}). The figure  shows at upper left  a cut through the solution at fixed $z=0.35$ which illustrates clearly a reversal in the azimuthal magnetic field.  At upper right  we give a three dimensional vector plot  of the same example  (but with larger $v$) that demonstrates the `X type' behaviour of the field. At lower right we  show the Faraday screen rotation measure (RM) showing the sign reversal or `parity inversion' in the first quadrant above the disc.  {The reddish to reddish -blue colour is generally positive, while the dark blue to green to yellow or orange is generally negative. Of course the signs in the same quadrant may be reversed as there is an arbitrary constant amplitude, but antisymmetry across the disc should hold.  Axial symmetry defines the other quadrants. }
Finally at lower left we show a typical field line  converging onto the axis of the galaxy. Once again these are for a specific, illustrative, choice of parameters that do not exhaust the possibilities. {\it The rotation measure should be antisymmetric across the plane }(i.e. to the fourth quadrant) according to the argument above. Our Unit is one disc radius, covered by $50$ grid Units. 

{In all sections of this paper the RM is calculated as the integral of the parallel magnetic field along the line of sight, for each line of sight, assuming that the electron density is constant. Thus it is a simple form of Faraday screen. The relevance to observations is certainly not adequate in this form, but improved calculations require a detailed model of the halo. This is beyond the scope of this work but not beyond future work based on these fields. In axial symmetry the result for an edge-on galaxy is  the integral of the azimuthal magnetic field along the line of sight. To the extent that the tangent point to a circle of radius $r$ has the strongest field component our RM is approximately the rotation measure produced near this point.} 
\section{Time dependent Scale Invariance}

It is relatively easy to introduce a scale invariant time dependence into the dynamo equations, which take the  basic form (the time derivative is omitted in the previous sections)
\be
\partial_t{\bf A}={\bf v}\wedge \nabla\wedge {\bf A}-\eta\nabla\wedge\nabla\wedge {\bf A}+\alpha_d\nabla\wedge {\bf A}.\label{eq:Afield}
\ee
The time dependence will simply be a power law in time or (in the limit of zero similarity class) an exponential  time factor. The geometry of the magnetic field remains `self-similar' over the time evolution, and we can therefore study the geometry without requiring a fixed epoch. We have however no way of bringing the dynamo into a steady state, although the `alpha effect', resistive diffusion and mean velocity field vary in a consistent fashion whether the dynamo grows or decays. It is important for what follows to observe that the time derivative  in equation (\ref{eq:Afield}) is taken at a fixed spatial point. We do not therefore differentiate the unit vectors.

The principal reason for introducing the time dependence is that it removes the over determined nature of the  equations for the vector potential. This problem was revealed in \cite{Hen2017} for the spatially axially symmetric; scale invariant, and steady, dynamo equations. The problem remains only if $a=2$, since then the equations reduce to those of the steady state. We will exclude this similarity class in this section. The  spatial magnetic field remains self-similar  to itself as time progresses, so that we do not have to specify a particular time when analyzing the spatial dependence. Nevertheless, each similarity class predicts a distinct time dependence which can be compared eventually with evolutionary time scales.

We proceed following a convenient  general technique (e.g. \citet{CH1991}, \citet{Hen2015}) for finding scale invariance, where now the time variable (rather than the radius) is the direction of the major scaling symmetry. We will in addition look for the same spatial conical geometry that we used in the preceding sections.

The temporal scale invariance allows us to write for the spatial and temporal dependence of the various  fields 
\bea
{\bf A}&=&\bar{\bf A}(R,Z)e^{(2-a)\delta T},\nonumber\\
{\bf b}&=& \bar{\bf b}(R,Z)e^{(1-a)\delta T}\equiv \nabla\wedge {\bf A},\nonumber\\
{\bf v}&=& \bar{\bf v}(R,Z)e^{(1-a)\delta T}.\label{eq:tempfields}
 \eea   
 The variables $\{R,Z,T\}$ are
 \bea
 R&=&re^{-\delta T},~~~~~~Z=ze^{-\delta T},\nonumber\\
 e^{\alpha T}&=& 1+\tilde\alpha_d\alpha t,\label{eq:tempvariables}
 \eea
 and the sub-scale helicity and diffusion parameters are given respectively by
 \bea
 \alpha_d&=&\bar\alpha_d(R,Z)e^{(1-a)\delta T},\nonumber\\
\eta&=& \bar \eta(R,Z) e^{(2-a)T},\label{eq:tempparams}
\eea
which determines their evolution in space and time.

The constant $\tilde\alpha_d$ will be defined below. The quantities $\delta$ and $\alpha$ may be thought of as reciprocal temporal scales for the spatial and temporal Dimensions respectively, wherever they occur in physical quantities. The choice of $a\equiv \alpha/\delta$ , gives the ratio of spatial power  to temporal power in the  Dimensions of some globally conserved quantity, which may not be otherwise defined. This is an advantage of this approach. The traditional argument from Dimensions alone requires assigning a global constant `a priori' . For example this global constant is $GM$ for Keplerian orbits so that $a=3/2$ (i.e. $a=3/2 $ from $[GM]=L^3/T^2$), but we do not know this generally in complicated systems. Leaving it as a parameter allows the significance of the choice to be studied without making an explicit choice (see Table~\ref{table:a} for more explicit examples). 

Making these substitutions into equations (\ref{eq:Afield}) removes the time dependence in favour of three partial differential equations in $R$ and $Z$. They are not physically complete without a  compatible prescription for the sub-scale helicity and the diffusion coefficients $\bar\alpha_d$ and $\bar\eta$. Once these are given the partial differential equations are quite general.  However we simplify these equations and reduce  them to ordinary equations through the following assumptions. We require the dependence on $R$ and $Z$ to take the form of a  unique dependence on $\zeta=Z/R=z/r$ just as in the steady state. In order for this to be a successful ans\"atz one is compelled to fix the sub-scale helicity and diffusion coefficients  as well as $\bar {\bf v}$ as follows 
\bea
\bar\alpha_d(R,Z)&=&\delta R~~\tilde\alpha_d,\nonumber\\
\bar \eta&=& \delta R^2~~\tilde\eta,\label{eq:tempspacescaling}\\
\bar {\bf v}(R,Z)&=&\delta R~~\tilde\alpha_d~~\{u,v,w\}.\nonumber
\eea

Moreover, after calculating $\nabla\wedge {\bf A} $, we find (recalling the second of equations (\ref{eq:tempfields}) for the complete magnetic field) that the magnetic field has the temporal scale free, mean field, form
\be
\bar {\bf b}=\frac{\tilde b(\zeta)}{R}\equiv \frac{1}{R}(-\bar A'_\phi,~~\bar A'_r+\zeta\bar A'_z,~~\bar A_\phi-\zeta\bar A'_\phi),\label{eq:btilde}
\ee
where the vector potential components are all functions of $\zeta$.

At this stage $\tilde\eta$ and $\tilde\alpha_d$ may still be arbitrary functions of $\zeta$, that is they may vary  from one cone to another. To avoid the recourse to poorly known physical conditions, we will proceed here with {the ratio} of these quantities constant. The dynamo equations to be solved are now reduced to the following 
\bea
(2-a)\bar A_r(\zeta)&=& -\bar A'_\phi\Delta+\frac{d}{d\zeta}( \bar A'_r+\zeta \bar A'_z)+v(\bar A_\phi-\zeta\bar A'_\phi)-w(\bar A'_r+\zeta\bar A'_z),\nonumber\\
(2-a)\bar A_\phi(\zeta)&=& (\bar A'_r+\zeta\bar A'_z)\Delta+(\bar A''_\phi(1+\zeta^2)+\zeta\bar A'_\phi-\bar A_\phi)+u(\zeta \bar A'_\phi-\bar A_\phi)-w\bar  A'_\phi,\nonumber\\
(2-a)\bar A_z(\zeta)&=& (\bar A_\phi-\zeta\bar A'_\phi)\Delta+\zeta\frac{d}{d\zeta}(\bar A'_r+\zeta \bar A'_z)+u(\bar A'_r+\zeta\bar A'_z)+v\bar A'_\phi.\label{eq:fullequs}
\eea

In these equations we have once again set the sub scale Reynolds number or `dynamo number'  (e.g. \citet{B2014}) equal to 
\be
\Delta=\frac{\tilde\alpha_d}{\tilde\eta}.\label{eq:dynamono}
\ee
For this  general form to be correct $\tilde\eta$ should take the place of $\tilde\alpha_d$ in the expression for $\bar{\bf v}$ (cf equation (\ref{eq:tempspacescaling})) and in the expression for the time variable (equation (\ref{eq:tempvariables})) because the general form assumes that there is a non-zero diffusion coefficient. Should this  term instead be zero, then the expressions become what we have given above  with $\tilde\alpha_d$ replacing $\tilde\eta$. {Moreover the second order terms in equations (\ref{eq:fullequs}) vanish and so $\tilde\alpha_d$ cancels out of the equations.  This renders formally $\Delta=1$ in that zero resistivity case. }

Equations (\ref{eq:fullequs}) may be reduced to a single fourth order equation for $\bar A_\phi$, or to two second order equations for $\bar A_\phi$ and $\tilde b_\phi$. The latter two equations for constant $\{u,v,w\}$ are 
\bea
\tilde b_\phi\Delta&=& (3-a+u)\bar A_\phi+(w-\zeta(1+u))\bar A'_\phi-(1+\zeta^2)\bar A''_\phi,\label{eq:phi component}\\
\bar A''_\phi(1+\zeta^2)\Delta&=& \tilde b''_\phi(1+\zeta^2)+\tilde b'_\phi(\zeta+\zeta u-w)-(2-a)\tilde b_\phi.\label{eq:r+z components}
\eea
One sees that the azimuthal velocity component $v$ does not appear. This may be due to the form of the velocity required in equation (\ref{eq:tempspacescaling}) where, in the pattern frame, the rotation is rigid. The exponential growth that appears in the limit that $\alpha=0$ must  nevertheless refer to a constant jump (there is some globally constant quantity with the Dimension of reciprocal time) in pattern speed relative to a boundary. 

These equations complete  the formulation of the mathematical problem , since the other magnetic field components follow from equation (\ref{eq:btilde}) in terms of $\bar A_\phi$.  There is a slight variation when we allow $w=u\zeta$, which permits an outflow to increase near the axis of the galaxy as $\zeta$ increases for constant $u$.  This requires setting $w=u\zeta$ in the previous equations and adding $-u\tilde b_\phi$ to the right hand side of equation (\ref{eq:r+z components}).

We observe that if $\Delta=0$ so that there is no turbulent dynamo,  then the equations decouple. Moreover in the constant velocity case with $u=-1$, the equations for $\tilde b_\phi$ and $\bar A_\phi$ are the same so that they can differ only by constant factors. This makes more sense in the case $a=\alpha=0$ since then there is presumably a global `dynamo' created by rotation with respect to an external boundary. The solution may be found in terms of hypergeometric functions, but we prefer to look at other cases here.  

One limit that can always be reduced to a linear second order equation is when the resistive diffusion is neglected so that the conductivity is infinite. Provided that some outflow is allowed, this still allows for realistic growth of the magnetic field  due to the sub-scale dynamo  by advection into  and through the halo.  Should there be inflow, the growth may be enhanced. In the limit of zero diffusion the governing equations become 
\bea
\tilde b_\phi&=&(2-a+u)\bar A_\phi+(w-\zeta u)\bar A'_\phi,\label{eq:phi-eta0-comp}\\
0&=& \bar A''_\phi((w-\zeta u)^2+1+\zeta^2)+2(2-a)(w-\zeta u)\bar A'_\phi+(2-a)(2-a+u)\bar A_\phi.\label{eq:r+z+phi})
\eea 
 One can verify directly from the equations for $\bar A_r$ and $\bar A_z$ in equations (\ref{eq:fullequs}--without the total derivative in the first and third equations and the bracketed middle term in the second equation) that $\bar b_\phi=\bar A'_r+\zeta \bar A'_z$ is consistent with equations (\ref{eq:phi-eta0-comp}) and (\ref{eq:r+z+phi}). However such a procedure relies numerically  on the accuracy of the solution to equation (\ref{eq:r+z+phi}). Were it not for the axial symmetry, a more secure way of guaranteeing zero divergence of the magnetic field would be to solve for $\bar A_r$ and $\bar A_\phi$ and then write $\bar b_\phi=\bar A'_r+\zeta \bar A'_z$. This must be done without axial symmetry as we shall see in a related paper, but in axial symmetry we can avoid this because  the azimuthal field component does not contribute to the divergence.
 
Once  equation (\ref{eq:r+z+phi}) is solved the other field components follow again from equations (\ref{eq:btilde}) (recalling equation (\ref{eq:phi-eta0-comp})).
The solution may always be found in terms of hypergeometric functions. There are several special cases however. One of them is to take $w=u\zeta$ with zero diffusion, which requires allowing $w$ to vary during the manipulations as in the preceding equations.  Equation (\ref{eq:phi-eta0-comp}) is valid on making the substitution  for $w$ directly, while equation (\ref{eq:r+z+phi}) requires making the substitution  for $w$ (so that the term in the first derivative vanishes) plus changing the factor multiplying $\bar A_\phi$  to $(2-a+u)^2$. In the following sub-sections we present various time dependent examples for specific choices of parameters in this infinite conductivity limit.

\subsection{The $w=u\zeta$ Zero diffusion Dynamo}

This example is independent of $v$ and allows a coupling between the radial and vertical velocities such that the vertical velocity increases near the axis of the galaxy. The parameters $a$ and $u$ are arbitrary. The growth rate is a power law $\propto (1+\tilde\alpha_d\alpha t)^{(2/a-1)}$.
The equation for $\bar A_\phi$ is 
\be
\bar A''_\phi(1+\zeta^2) +(2-a+u)^2\bar A_\phi=0,\label{eq:tempw=uzeta}
\ee
for which the solution is
\bea
\bar A_\phi&=&C1(1+\zeta^2)F(3/4+(i/8)rad(a,u),3/4-(i/8)rad(a,u);1/2;-\zeta^2)\nonumber\\
&+&C2(1+\zeta^2)\zeta F(5/4+(i/8)rad(a,u),5/4-(i/8)rad(a,u);3/2;-\zeta^2),\label{eq:soltemp1}
\eea
where
\be
rad(a,u)\equiv \sqrt{(4a-4u-6)}\sqrt{(4a-4u-10)},
\ee
and $F(a,b;c;z)$ is the hypergeometric function. 
The other components of the magnetic field now follow from equation (\ref{eq:phi-eta0-comp}) and equations (\ref{eq:btilde}).

\begin{figure}
\begin{tabular}{cc} 
\rotatebox{0}{\scalebox{0.5} 
{\includegraphics{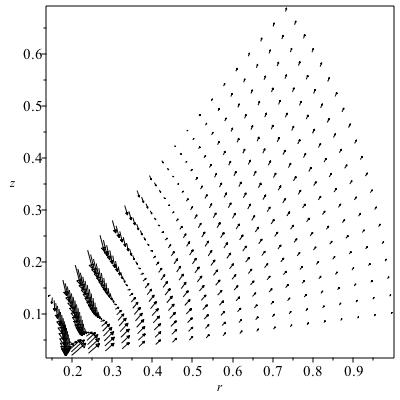}}}&
\rotatebox{0}{\scalebox{0.5} 
{\includegraphics{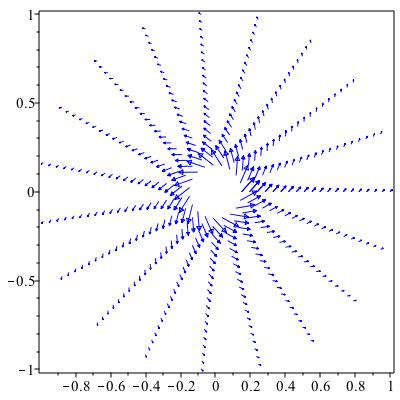}}}\\
{\rotatebox{0}{\scalebox{0.5} 
{\includegraphics{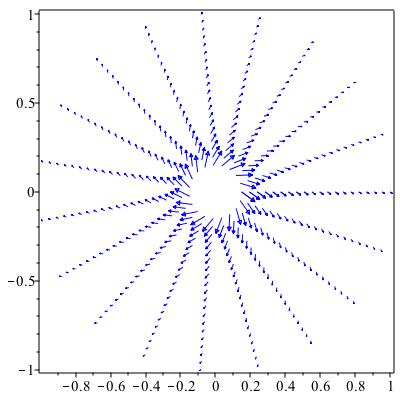}}}}&
\rotatebox{0}{\scalebox{0.55} 
{\includegraphics{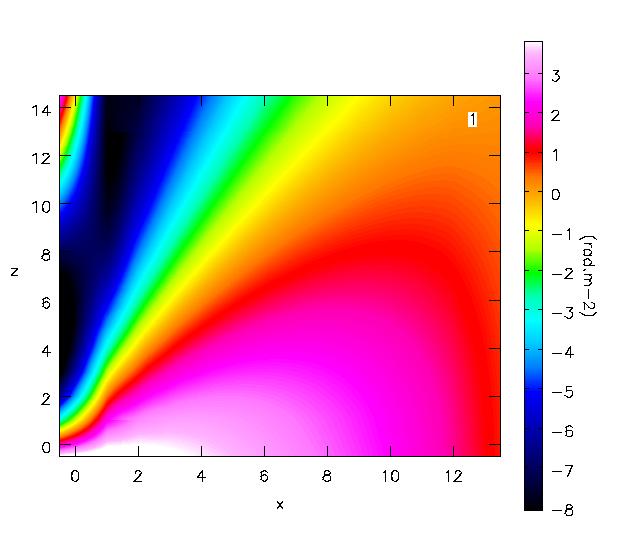}}}
\end{tabular}
\caption{ The parameter set is $\{ a,u,r,z,C1,C2\}$ for this example. At upper left a  poloidal cut through the polar axis for the parameter set $\{1,2,r,0.25,1,0\}$ is shown with $0.2\le r\le 1$ and $0.1\le z\le 0.75$. At upper right and at lower left the magnetic field on the conical surfaces $z=0.25r$ and $z=0.75r$ are shown for the same parameter set and range in $r$.  At lower right we show the RM screen in the first quadrant ( but with $u=0.5$) where the disc radius and halo height  both run over $[1,~13.6]$ in grid Units ($[1/15,0.9]$ in Units of the disc radius).}    
\label{fig:tempw=uzeta}
\end{figure}

This solution has been illustrated 
 in figure (\ref{fig:tempw=uzeta}).  {\it The 'parity change', 'X type' structure and `diagonal RM' are evident in this example}. A typical magnetic field line winds off-axis until it reaches a height where the winding becomes straight. We include several cuts in figure (\ref{fig:tempw=uzeta}) that illustrate the magnetic field structure.

 We illustrate only the class $a=1$ in this section although any $a$ except $a=2$ is allowed. The various cuts in figure (\ref{fig:tempw=uzeta}) together with the RM screen show clearly the sign reversal in the azimuthal magnetic field along the diagonal $z\approx 0.5 r$. The axially symmetric spiral structure is apparent. {The poloidal cut at upper left illustrates the `X-type' magnetic field in this plane.  This behaviour occurs at small angle to the plane ($\zeta<1$) except at large radius where the field weakens and turns parallel to the axis}.  The class $a=1$ allows for a globally constant velocity, such as that of the disc at a fixed height. This model has an AGN type outflow along the minor axis of the galaxy as both the magnetic field and the outflow increase there. Larger outflows tend  in general to reduce the field strength near the disc.

\subsection{ The $a=u=0$  Zero Diffusion Dynamo}

This example is also independent of $v$ although $a=0$ implies some abrupt jump in the rotation velocity. The time dependence is exponential.
In this case equation (\ref{eq:r+z+phi}) becomes 
\be
(1+\zeta^2+w^2)\bar A''_\phi+4w\bar A'_\phi+4\bar A_\phi=0.\label{eq:a=u=0}
\ee
The solution can also be found in terms of hypergeometric functions of somewhat more complicated arguments than in the preceding example. 

\begin{figure}
\begin{tabular}{cc} 
\rotatebox{0}{\scalebox{0.5} 
{\includegraphics{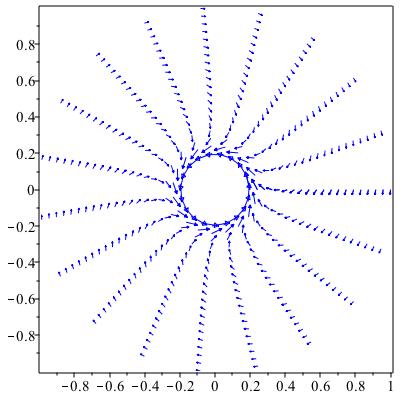}}}&
\rotatebox{0}{\scalebox{0.5} 
{\includegraphics{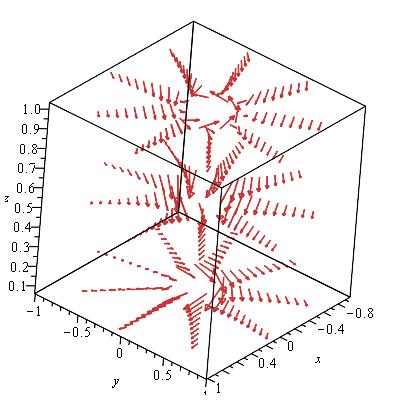}}}\\
{\rotatebox{0}{\scalebox{0.5} 
{\includegraphics{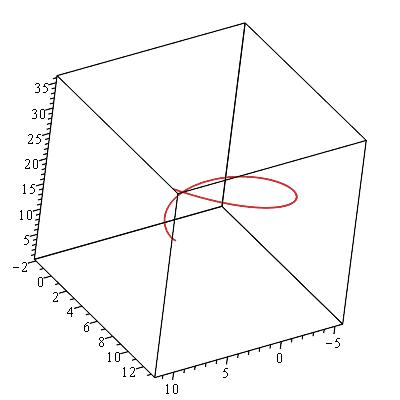}}}}&
\rotatebox{0}{\scalebox{0.57} 
{\includegraphics{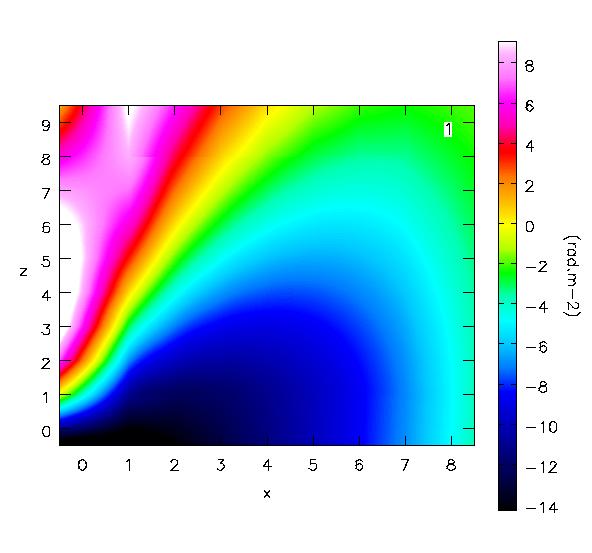}}}
\end{tabular}
\caption{ We have the parameterspace $\{ w,r,z,C1,C2\}$  for this figure. At upper left we show a 2d cut with $\{0.5,r,0.6,1,0\}$ with $0.2\le r\le 1$.  At upper right the magnetic field vectors in 3d are shown over $0.25\le r\le 1$ and $0.1\le z\le 1$ for the same parameters. At lower left a fieldline starting at $\{r,\phi,z\}=\{0.15,0,0.01\}$ is shown for the same parameters. The spiral is `off axis' in that its axis does not coincide with that of the galaxy. Such spirals begin at each initial value of $\phi$. At lower right we show the RM screen  over $\{0.1\le r\le 0.9\}$ and $\{0.1\le z\le 1$ (in Unts of disc radii)  in the first quadrant for the same parameters. }
\label{fig:tempcase3}
\end{figure}

We show some examples in figure(\ref{fig:tempcase3}). The field has even symmetry in the azimuthal magnetic field but antisymmetry for the radial component across the disc . The magnetic field behaves this way when $w$ reverses sign as it should if  the flow does not represent a wind.  
Both cases of $\{C1,C2\}$, namely $\{1,0\}$ and $\{0,1\}$  and hence $\{1,1\}$ satisfy these disc conditions. One can not change the sign of $C1$ or $C2$ since the vertical field must be continuous across the disc.

 The cut at upper left in this figure shows  nicely the sign reversal in the azimuthal magnetic field in radius. At upper right the three dimensional magnetic vectors are illustrated. The field is nearly `X-type near the plane but becomes more azimuthal at height above the disc. At lower left we show a section of a
 typical field line in the first quadrant. It would continue into the fourth quadrant with a slight discontinuity due to the change in direction of $B_r$ on crossing the disc. The spiral is `off axis' relative to the minor axis of the galaxy.  At lower right we show the RM screen in the first quadrant for this case. It corresponds to the three dimensional structure in the panel at upper right. The sign changes sign approximately along a diagonal.

\subsection{The $a=1$, $u=0$ Zero Diffusion Dynamo}

\begin{figure}
\begin{tabular}{cc} 
\rotatebox{0}{\scalebox{0.6} 
{\includegraphics{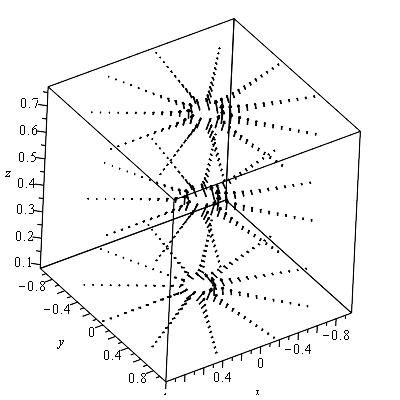}}}&
\rotatebox{0}{\scalebox{0.6} 
{\includegraphics{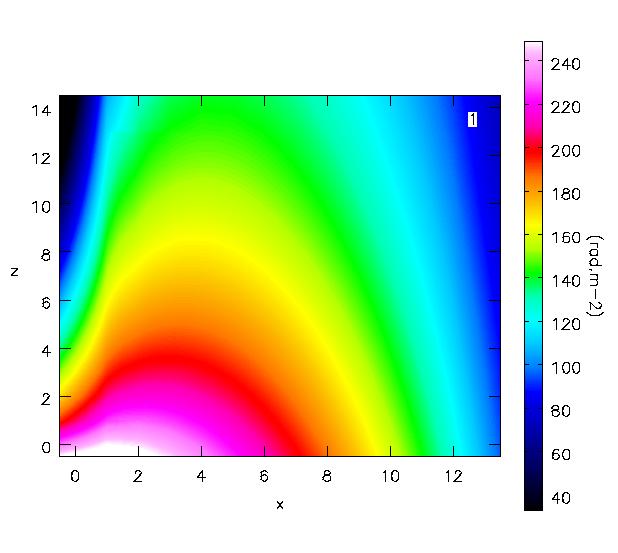}}}
\end{tabular}
\caption{ The figure on the left has the parameter set $\{w,r,z,C1,C2\}=\{2,r,z,0,1\}$ and $0.15\le r\le 1$, $0.1\le z\le 0.75$. The field shows little variation. At right we show the corresponding RM screen. There is no sign change but the red end of the colour spectrum is somewhat stronger in amplitude. }    
\label{fig:tempcase4}
\end{figure}

We include this case mainly as an example where there is no sign change in the azimuthal magnetic field. The calculation is 
illustrated in figure (\ref{fig:tempcase4}). We also state explicitly in this section the boundary conditions that are required on crossing the disc, although they have also been met in our previous examples.

These require generally that 
\be
b_z(w,0^+)=b_z(-w,0^-),
\ee
and one of 
\be
b_{\phi}(w,0^+)=\pm b_{\phi}(-w,0^-),~~~~~b_r(w,0^+)=\pm b_r(-w,0^-).
\ee
Moreover, if the disc sub-scale dynamo is to have similar effect on both sides, then these conditions should also hold well away from the plane. In the present example, this is achieved when the second form of the solution is used (i.e. $(C1,C2)=C2(0,1)$). The azimuthal field is then symmetric across the disc, while the radial magnetic field is asymmetric.. The other possible solution  would be required to describe {\it global} asymmetry across the disc (i.e. not the boundary condition at the disc), such as might be required to describe a wind through the disc from the environment.

   
   The equation to be solved in this case is (\ref{eq:r+z+phi}) 
   \be
   \bar A''_\phi(1+w^2+\zeta^2)+2w\bar A'_\phi+\bar A_\phi=0,\label{eq:Aphia1u0}
   \ee
   after which equations (\ref{eq:btilde}) and (\ref{eq:phi-eta0-comp}) yield the magnetic field.
   
  Figure (\ref{fig:tempcase4}) shows the three dimensional structure and the RM screen for one example. It is a case where the poloidal magnetic field is rather dipolar (presumable closing well outside the disc at $r=1$) and there is only a significant azimuthal field near the plane. The latter does not change sign and the strength of the RM screen varies relatively slowly being less than a factor of three over most of the image.

\section{Discussion and Conclusions}
Our objective in this paper was to make contact with the previous scale-invariant formulation \cite{Hen2017} and to extend that formulation to self-similarity in time.  It was also to confirm the previous approximate work by a selection of exact solutions. Our scale invariant approach is independent of the many previous theoretical contributions based on more physical considerations, which would require a full review paper to fairly acknowledge. A reader may pursue this question on his own  by addressing \cite{KF2015},  \cite{M2015}, \cite{Black2015}, \cite{B2014} ,and \cite{FT2014}. 

The scale invariant approach avoids the various physical arguments regarding the sub scale helicity, resistive diffusion, and magnetohydrodynamics by requiring all of these quantities to be compatible with the assumed symmetry. This is to be sure a somewhat simplified approach, but we think that it is justified here by the sample results given. It is worth summarizing however the important assumptions that we have made in addition to the scale invariance or self-similarity. 

{  The  application of self-similar symmetry to axially symmetric dynamo theory requires in the steady state only a function of $\zeta$ to be found. The symmetry assumption determines the explicit radial dependence once the `class' is assigned. When time dependence is introduced, only the dependence in time is initially fixed by the similarity `class'. This means that our problem is reduced to partial differential equations in the variables $R$ and $Z$. At this stage  the various auxiliary quantities such as diffusivity, helicity and velocity may also be a function of these variables. However this does not advance us much beyond standard numerical approaches, given that we seek analytic examples.

For this reason we have introduced additional assumptions in the time dependent case. The additional assumption of spatial scale invariance, inspired by the steady state scale invariance, reduces the $\{R,Z\}$ dependence to a dependence on $\zeta$.  This succeeds only by imposing the conditions (\ref{eq:tempspacescaling}), which define an explicit radial and temporal dependence. These  conditions still allow the auxiliary quantities to vary on cones (in both steady and non steady dynamos) but the imposed radial and temporal dependence remains. However we have taken the ratio of helicity to diffusivity to be constant for simplicity (and  the pattern velocity to be proportional to the diffusivity), which warrants some justification.  

Physically we might expect  the helicity, the diffusivity, and the velocity to decrease on leaving the galactic disc. \footnote{ Conceivably they might rise again on approaching the galactic axis at large $\zeta$.} Individually, the helicity and diffusivity and velocity are free to do this in our formulation, but  the constant ratios may appear arbitrary. If both the helicity and the diffusivity are due to sub scale `turbulent' motions, then it is not unlikely that they vary together on cones. However this also requires the pattern velocity to scale in the same way as the sub scale motion, which is compatible with pure scale invariance.  However future work must determine whether all this can be reconciled with dynamics of turbulent halo flow.
 
 There is some preliminary observational evidence that we may not have restricted the physics unduly. }
 For example, 
  {\it without further assumptions}, our examples demonstrate a) the presence of magnetic spirals (here axially symmetric) \cite{Beck2015,fri16}, and b) `X-type fields' that have been observed frequently \cite{Beck2015}, \cite{Kr2015} and indeed throughout the CHANG-ES sample \cite{WI2015}.  Moreover it {\it predicts} \cite{Hen2017}, \cite{Hen2017b} {\it c) the remarkable rotation measure structure (reversing signs) now discovered in the halo of NGC~4631 \cite{CMP2016} even in axial symmetry}. This has appeared in previous work under the heading of `parity inversion' (e.g. \cite{MS2008}), but it seemed to be controversial in axial symmetry.  We now confirm our previous approximate results by finding similar behaviour in exact solutions. We finally {\it predict} that the planar magnetic spirals observed in face-on galaxies \cite{Beck2015} are `lifted' into the galactic halo.

 To elaborate further on the specific example of X-shaped fields, a
 key result of this paper is that classical dynamo theory alone, as demonstrated here, can explain a variety of observational properties {\it without invoking additional mechanisms}. For example, it has been suggested that the X-shaped fields observed in NGC~253 \citep{hee09} may require dynamo action {\it plus} a strong nuclear wind. Yet galaxies without super winds can also show X-shaped fields (same reference).  Here, we suggest that the X-shaped fields are present first and that the wind will result from the underlying star formation rate plus the fact that X-ray shaped fields may help to supply the appropriate topology for cosmic ray particles to escape.  The `illumination' of the existing X-shaped fields depends on available escaping particles. 
{ The X-type structure is of course generally reminiscent of a kind of `conical symmetry'. Moreover a recent discovery \cite{MIW2017} of the proportionality between radio scale height and radio (radial) scale length is at least compatible with a conical emission structure in each galaxy.}

We have not tried here  to catalogue all the possible behaviours that may be discovered with other choices of parameters.{ We believe that the observational characteristics discussed above are generic.} Extraordinary behaviour should be accessible to other workers, particularly when fitting observations.  This is expected to be the next phase of this project. As a convenience, we will make  available the MAPLE\footnote{{\bf {\tt maplesoft.com}}} scripts that we have used to calculate these results. 

{ One  parameter that is left deliberately undetermined is the class $a$, which is shown in our table to be related to a globally conserved constant. In the case of the steady state $a=1$ would seem to account for the constant rotation velocity.  With time dependence however, we introduce a multitude of effects  that might occur during the formation of the galaxy and its magnetic field. We have given the physical interpretation of most of the possibilities (not exhaustive) in the table, and it is not clear to what extent angular momentum transfer, or flux conservation, or even Keplerian orbits of accreting galaxies, may play a r\^ole. For this reason we have encouraged the attitude that it is a parameter to be used in fitting data.    }

It is always possible, as is frequently observed in complex interacting systems (\citet{Barenblatt96}, \citet{Hen2015}), that a galactic disc-halo is asymptotically scale invariant in the mean. { In the case of a spiral galaxy `asymptotic'  applies to the disc outside the bulge and to the halo below any boundary with the intergalactic medium. If the halo field is linked to the intergalactic field (e.g. \cite{HI2016}), then the asymptotic scale invariance would have to imply the whole structure.}

Ultimately the justification for this  non-dynamical, scale invariant, approach is that {\it it allows a relatively rapid exploration of the effects of different flow geometries and global constants}. We have illustrated this both for a steady state and for self-similar time dependence. However much is left to be done in exploring parameter space in conjunction with data.  Whether this is a correct description must be left to the observations.

\vskip 0.5truein

\noindent{\bf Acknowledgements}\\
This work has been supported by a Discovery Grant to JI by the Natural Sciences and Engineering Research
Council of Canada and a Reinhardt Fellowship to AW from Queen's University. An anonymous referee is to be thanked for constructive criticism and helpful remarks.


      \label{lastpage}
\end{document}